\documentclass[review,fleqn]{elsarticle}

\usepackage{amsmath}
\usepackage{amsfonts}
\usepackage{amssymb}

\usepackage{graphicx}
\usepackage{nccmath}
\usepackage{float}
\usepackage{caption} 
\usepackage{titlesec}
\titlespacing{\section}{0pt}{2ex}{0ex}

\biboptions{sort&compress}

\begin{document}
\setlength{\abovedisplayskip}{5pt}
\setlength{\belowdisplayskip}{5pt}
\setlength{\abovedisplayshortskip}{0pt}
\setlength{\belowdisplayshortskip}{0pt}

\begin{frontmatter}
\title{Pseudospin and spin symmetry in the relativistic generalized Woods-Saxon potential including Coulomb-like tensor potential}

\author{J. Akbar}
\author{A. Suparmi\corref{mycorrespondingauthor}}
\cortext[mycorrespondingauthor]{Corresponding author}
\ead{soeparmi@staff.uns.ac.id}
\author{C. Cari}

\address{Department of Physics, Sebelas Maret University, Surakarta, Indonesia}

\begin{abstract}
The Dirac Equation is solved approximately for relativistic generalized Woods-Saxon potential including Coulomb-like tensor potential in exact pseudospin and spin symmetry limits. The bound states energy eigenvalues are found by using wavefunction boundary conditions, and corresponding radial wavefunctions are obtained in terms of hypergeometric function. Some numerical examples are given for the dependence of bound states energy eigenvalues on quantum numbers and potential parameters.\\
\end{abstract}

\begin{keyword}
Pseudospin symmetry; Spin symmetry; Dirac Equation; 
\\ 
\hspace{5em} Woods-Saxon potential; Tensor potential
\end{keyword}
\end{frontmatter}

\section{Introduction}
The pseudospin symmetry concept was introduced about 40 years ago through the observation of quasi-degeneracy in some nuclei between single nucleon state with quantum numbers ($n, l, j = l + 1/2$) and ($n-1, l + 2, j = l + 3/2$), where $n$, $l$ and $j$ are the radial, the orbital, and the total angular momentum quantum numbers, respectively. These two states are known as pseudospin doublets and have the same "pseudo" orbital angular momentum quantum number, $\tilde{l} = l + 1$, and pseudospin quantum number $\tilde{s} = 1/2$. The exact pseudospin symmetry occurs when doublets with $j = \tilde{l} \pm \tilde{s}$ are degenerate \cite{arima1969pseudo, hecht1969generalized}. This doublet structure has been successfully explained the nuclear physics phenomena, including deformation, superdeformation, identical bands, and magnetic moments \cite{bohr1982pseudospin, dudek1987abundance, troltenier1994validity, stuchbery1999magnetic, stuchbery2002magnetic, stephens1990pseudospin}. Meanwhile, the spin symmetry concept is used to explain the absence of spin-orbital splitting between a single nucleon state with opposite spin ($n, l, j = l \pm 1/2$). By using spin symmetry concepts we can apply the triaxial, axially deformed, and spherical oscillator potentials to the eigenvalues of an antinucleon embedded in a nucleus \cite{ginocchio2004relativistic}, and it has been shown \cite{zhou2003spin} that the spin symmetry in the antinucleon spectra is much better developed than the pseudospin symmetry in normal nuclear single-particle spectra. Pseudospin symmetry occurs when $\mathit{\Sigma}(r) = S(r) + V(r) =$ constant, and spin symmetry occurs when $\mathit{\Delta(r)} = S(r) - V(r) =$ constant, where $S(r)$ is attractive scalar potential and $V(r)$ is repulsive vector potential. The pseudospin and spin symmetry are exact under conditions $d\mathit{\Sigma}(r)/dr = 0$ and $d\mathit{\Delta}(r)/dr = 0$, respectively \cite{ginocchio1997pseudospin, meng1998pseudospin, meng1999pseudospin}. Since that time, the Dirac equation's analytical solutions have been investigated by many papers for solvable potentials, such as Woods-Saxon potential \cite{candemir2014bound, xu2006pseudospin, aydougdu2010pseudospin, alberto2002pseudospin, lisboa2010spin}, Pseudoharmonic potential \cite{aydougdu2010exact}, Mie-Type potential \cite{hamzavi2010exact}, Morse potential \cite{berkdemir2006pseudospin}, Eckart potential \cite{soylu2008kappa}, and Asymmetrical Hartmann potential \cite{guo2007exact}, by different methods.

Tensor coupling under the conditions pseudospin and spin symmetry has been studied in references \cite{akcay2009dirac, akcay2009exact, alberto2005tensor, lisboa2004pseudospin}. They have found that tensor interaction can remove the degeneracy between two states in pseudospin and spin doublets. On the other hand, the tensor coupling has been applied to the nucleon in effective chiral lagrangians for nuclei \cite{furnstahl1998nuclear} and describing the bound states of both nucleons and antinucleons in the relativistic Hartree approach \cite{mao2003effect}. Thus, it is very convenient to use nuclear potentials like the generalized Woods-Saxon potential and tensor potential for investigating the energy eigenvalues of nuclei and their corresponding eigenfunction. The generalized Woods-Saxon potential consists of Woods-Saxon plus Woods-Saxon surface potential. This surface term might be a useful model for describing the surface interactions between nucleons. The solution of $s$-wave Dirac equation for Woods-Saxon potential has been solved by Guo et al. \cite{guo2005solution} and is discussed in \cite{bila2006comment, guo2006reply}, they found that the condition of existing bound state energy only exist for $V+S<0$ or $C_{ps}<0$ in the limit of exact pseudospin symmetry. Pahlavani et al. studied the the nuclear bound states using mean-field Woods-Saxon and spin-orbit potentials by using Nikiforov-Uvarof (NU) method, but they use this method carefully with considering wave function boundary conditions to get the energy eigenvalues equations and corresponding eigenfunctions \cite{pahlavani2012study}. Furthermore, Candemir et al. have solved the Dirac equation for the generalized Woods-Saxon potential and showed that the energy eigenvalues equation that provided by NU method is incorrect because the NU method does not take into account behavior of the wave function in vicinity $r=R$ \cite{candemir2014bound}. In the last three references, they carefully examine the asymtotic behavior of the wave function in vicinity $r=R$ using some relations of the hypergeometric function. Our purpose here is to go further and investigate the solution of Dirac equation for generalized Woods-Saxon potential including Coulomb-like tensor potential by using this wave function boundary conditions.

The present work is organized as follows: In Section 2, we give theoretical formalism for the Dirac equation. In Section 3 and 4, we solved and analyzed the Dirac equation's solution for generalized Woods-Saxon potential including Coulomb-like tensor potential in the limit of exact spin and pseudospin symmetry. The conclusions are summarized in Section 5.
\\

\section{Dirac Equation}

The general form of Dirac equation for fermionic massive spin$-1/2$ particles moving in a repulsive vector potential $V(\vec{r})$ and an attractive scalar potential $S(\vec{r})$ can be written as $(c=\hbar=1)$

\begin{eqnarray}\label{eq1}
\bigl( \vec{\alpha} \cdot \vec{p} +\beta(M+S)+V-i\beta \vec{\alpha} \cdot \hat{r} U(r) \bigr)\psi = \mathcal{E} \psi.
\end{eqnarray}
For spherical nuclei, the total angular momentum operator $\vec{J}$ and the spin-orbit coupling operator $\vec{\mathcal{K}}$ commute with the Dirac Hamiltonian, where $\beta, \vec{\sigma}$ and $\vec{L}$ are the Dirac matrix, Pauli matrix, and orbital angular momentum, respectively. The eigenvalues of the spin-orbit coupling operator are $\kappa=l>0$ and $\kappa=-(l+1)<0$ for unaligned spin $(j=l-1/2)$ and aligned spin $(j=l+1/2)$, respectively. Based on their angular momentum $j$ and $\kappa$, the wavefunctions can be written in the following form:

\begin{eqnarray}\label{eq2}
\psi_{n\kappa} \left(\vec{r}\right) =
\begin{pmatrix}
f_{n\kappa}\\
g_{n\kappa}\\
\end{pmatrix}
=\frac{1}{r}
\begin{pmatrix}
F_{n\kappa}(r)Y_{jm}^{l}(\theta ,\phi)\\
iG_{n\kappa}(r)Y_{jm}^{\tilde{l}}(\theta ,\phi)\\ 
\end{pmatrix}
,
\end{eqnarray}
where $n$ is the radial quantum number, and $m$ is the projection of angular momentum on the third axis. The angular part can be splitting off and leaving the upper and lower radial wavefunctions $F_{n\kappa}(r)$ and $G_{n\kappa}(r)$ as

\begin{eqnarray}\label{eq3}
\left( \frac{d}{dr}+\frac{\kappa}{r}-U(r)  \right) F_{n\kappa}(r)=\bigl(M+\mathcal{E}-\mathit{\Delta(r)} \bigr)G_{n\kappa}(r),  
\end{eqnarray}

\begin{eqnarray}\label{eq4}
\left( \frac{d}{dr}-\frac{\kappa}{r}+U(r)  \right) G_{n\kappa}(r)=\bigl(M-\mathcal{E}+ \mathit{\Sigma(r)} \bigr)F_{n\kappa}(r),
\end{eqnarray}
where $\mathit{\Delta}(r)=V(r)-S(r)$ and $\mathit{\Sigma}(r)=V(r)+S(r)$. Then, by eliminating $G_{n\kappa}(r)$ in Eq.(3) and $F_{n\kappa}(r)$ in Eq.(4), one can obtain Schrödinger-like equations for the upper and the lower components of the radial wavefunctions

\begin{equation}\label{eq5}
\begin{split}
&\left[ \vphantom{\frac{\frac{d\mathit{\Delta(r)}}{dr}\left(\frac{d}{dr}+\frac{\kappa}{r}-U(r) \right)}{M+\mathcal{E}-\mathit{\Delta(r)}}} 
       \frac{d^{2}}{dr^{2}} -\frac{\kappa(\kappa+1)}{r^{2}}+\left(\frac{2\kappa}{r}-U(r)-\frac{d}{dr} \right)U(r) \right. \\ 
& \qquad \ \left.
+\ \frac{\frac{d\mathit{\Delta(r)}}{dr}\left(\frac{d}{dr}+\frac{\kappa}{r}-U(r) \right)}{M+\mathcal{E}-\mathit{\Delta(r)}}\right] F_{n\kappa}(r)\\
& \qquad \ + \bigl(\mathcal{E}+M-\mathit{\Delta(r)} \bigr) \bigl(\mathcal{E}-M-\mathit{\Sigma(r)}\bigr)F_{n\kappa}(r)=0, 
\end{split}
\end{equation}

\begin{equation}\label{eq6}
\begin{split}
&\left[ \vphantom{\frac{\frac{d\mathit{\Sigma(r)}}{dr}\left(\frac{d}                        
{dr}-\frac{\kappa}{r}+U(r) \right)}{M-\mathcal{E}+\mathit{\Sigma(r)}}}
       \frac{d^{2}}{dr^{2}} -\frac{\kappa(\kappa-1)}{r^{2}}+\left(\frac{2\kappa}{r}-U(r)+\frac{d}{dr} \right)U(r) \right. \\ 
& \qquad \ \left.
-\ \frac{\frac{d\mathit{\Sigma(r)}}{dr}\left(\frac{d}{dr}-\frac{\kappa}{r}+U(r) \right)}{M-\mathcal{E}+\mathit{\Sigma(r)}}\right] G_{n\kappa}(r)\\
& \qquad \ + \bigl(\mathcal{E}+M-\mathit{\Delta(r)} \bigr) \bigl(\mathcal{E}-M-\mathit{\Sigma(r)}\bigr)G_{n\kappa}(r)=0, 
\end{split}
\end{equation}

\noindent
The solutions of Eq.(5) can be obtained with $\kappa(\kappa+1)=l(l+1)$ and its eigenvalues $\mathcal{E}_{n\kappa}=\mathcal{E}(n,l(l+1))$ depend only on $n$ and $l$. In the exact of spin symmetry, the states with $j=l \pm 1/2$ are degenerate for $l \neq 0$. Meanwhile, the solutions of Eq.(6) can also be obtained with $\kappa(\kappa-1)=\tilde{l}(\tilde{l}+1)$ and its eigenvalues $\mathcal{E}_{n\kappa}=\mathcal{E}(n,\tilde{l}(\tilde{l}+1))$ depend only on $n$ and $\tilde{l}$. By using condition in the exact of pseudospin symmetry, the states with $j=\tilde{l}\pm 1/2$ are degenerate for $\tilde{l}\neq 0$.
\\

\section{The pseudospin symmetry}
In the case of exact pseudospin symmetry ($d\mathit{\Sigma}(r)/dr=0,$ i.e., $\mathit{\Sigma}(r)=C_{ps}=$ const), we can assume $\mathit{\Delta}(r)$ and $U(r)$ are potential of the generalized Woods-Saxon type and potential of the Coulomb-like tensor, respectively, defined as follow

\begin{eqnarray}\label{eq7}
\mathit{\Sigma(r)}=C_{ps},\hspace{2em} \mathit{\Delta(r)}=-\frac{\mathit{\Delta}_0}{1+e^{\frac{r-R}{a}}}-\frac{\mathit{\Delta}_1 e^{\frac{r-R}{a}}}{\left(1+e^{\frac{r-R}{a}} \right)^2 }, \hspace{2em}  U(r)=-\frac{H}{r},
\end{eqnarray}
where $\mathit{\Delta}_0$ and $\mathit{\Delta}_1$ represent the depth of the potential well. The parameters $a$, $R$, and $H$ are the diffusivity of nuclear surface, the width of the potential well, and constants, respectively. On the other hand, Eq.(6) becomes

\begin{equation}\label{eq8}
\begin{split}
& \left[ \vphantom{\left(\mathcal{E}+M+\frac{\mathit{\Delta}_0}{1+e^{\frac{r-R}{a}}}+\frac{\mathit{\Delta}_1 e^{\frac{r-R}{a}}}{\left(1+e^{\frac{r-R}{a}} \right)^2}    \right)}
      \frac{d^2}{dr^2} - \frac{(\kappa+H)(\kappa+H-1)}{r^2} \right. \\
& \left. +\left(\mathcal{E}-M-C_{ps}\right) \left(\mathcal{E}+M+\frac{\mathit{\Delta}_0}{1+e^{\frac{r-R}{a}}}+\frac{\mathit{\Delta}_1 e^{\frac{r-R}{a}}}{\left(1+e^{\frac{r-R}{a}} \right)^2}    \right)     \right]G_{n\kappa}(r)=0.
\end{split}
\end{equation}

\noindent
There is no analytical solution of Eq.(8) with this potential model for $(\kappa+H)(\kappa+H-1)\neq 0$ because the pseudospin-orbit coupling term $(\kappa+H)(\kappa+H-1)/r^2$. Therefore, we shall use the Pekeris approximation \cite{pekeris1934rotation} by introducing the notations

\begin{eqnarray}\label{eq9}
x=\frac{r-R}{R}, \hspace{2em} \nu=\frac{R}{a},
\end{eqnarray}
and expanding the centrifugal potential in a series around $x=0 \ (r\approx R)$ as

\begin{eqnarray}\label{eq10}
V_{so}(x)=\frac{(\kappa+H)(\kappa+H-1)}{R^2}\left(1+x\right)^{-2} = \tilde{\gamma}\left(1-2x+3x^2+\cdots \right).  
\end{eqnarray}
where $\tilde{\gamma}=(\kappa+H)(\kappa+H-1)/R^2$. According to the Pekeris approximation, we shall replace the potential $V_{so}(x)$ with expression

\begin{eqnarray}\label{eq11}
V_{so}^{*}=\tilde{\gamma} \left(D_0+\frac{D_1}{1+e^{\nu x}} + \frac{D_2}{ \left( 1+e^{\nu x} \right)^2} \right),
\end{eqnarray}
where $D_0$, $D_1$, and $D_2$ are constants. We also expand this new form of centrifugal potential $V^{*}(x)$ in a series around the point $x=0\ (r\approx R)$

\begin{eqnarray}\label{eq12}
V_{so}^{*}=\tilde{\gamma}  \left[\left(D_0+\frac{D_1}{2}+\frac{D_2}{4} \right) - \frac{\nu}{4} \left(D_1+D_2 \right)x + \frac{D_2 \nu^2}{16}x^2+ \cdots  \right].
\end{eqnarray}
Comparing the equal powers of Eqs.(10) and (12), we get

\begin{eqnarray}\label{eq13}
D_0=1 - \frac{4}{\nu} + \frac{12}{\nu^2}, \hspace{2em}  D_1=\frac{8}{\nu} - \frac{48}{\nu^2}, \hspace{2em}  D_2=\frac{48}{\nu^2},
\end{eqnarray}
Now, we introduce a new variable $z=\left (1+e^\frac{r-R}{a}\right )^{-1}$ to derive the solution of Eq.(8). Then Eq.(8) becomes

\begin{fleqn}
\begin{equation}\label{eq14}
\begin{split}
& \left[\frac{d^2}{dz^2} + \frac{(1-2z)}{z(1-z)}\frac{d}{dz} 
+ \frac{1}{z^2(1-z)^2} \bigl(-\left(\tilde{\gamma}D_2 + \tilde{\mu} \right)a^2z^2 + \left(\tilde{\varpi}+\tilde{\mu}-\tilde{\gamma}D_1 \right)a^2z 
\right. \\ 
& \left.
\hspace{14.5em} \ - \left(\tilde{\gamma}D_0-\mathit{\tilde{\Lambda}} \right)a^2    
\vphantom{\frac{d^2}{dz^2}}  
\bigr)   
 \right]  G_{n\kappa}(z)=0.
\end{split}
\end{equation}
\end{fleqn}
where

\begin{eqnarray}\label{eq15}
\tilde{\varpi}=\left(\mathcal{E}-M-C_{ps}\right)\mathit{\Delta}_0, 
\end{eqnarray}

\begin{eqnarray}\label{eq16}
\mathit{\tilde{\Lambda}}=(\mathcal{E}+M)\left(\mathcal{E}-M-C_{ps} \right),
\end{eqnarray}

\begin{eqnarray}\label{eq17}
\tilde{\mu}=(\mathcal{E}-M-C_{ps})\mathit{\Delta}_1,
\end{eqnarray}
Eq.(14) can be reduced to a hypergeometric equation by setting the following factorization

\begin{eqnarray}\label{eq18}
G_{n\kappa}(z)=z^{\tilde{\sigma}}(1-z)^{\tilde{\tau}}w(z)
\end{eqnarray}
with

\begin{eqnarray}\label{eq19}
\tilde{\sigma}=\sqrt{\left(\tilde{\gamma}D_0-\mathit{\tilde{\Lambda}}  \right)a^2},
\end{eqnarray}

\begin{eqnarray}\label{eq20}
\mathit{\tilde{\chi}}=\sqrt{\bigl(\tilde{\varpi} - \tilde{\gamma}(D_1+D_2) \bigr)a^2},
\end{eqnarray}

\begin{eqnarray}\label{eq21}
\tilde{\tau}=\sqrt{\tilde{\sigma}^2-\mathit{\tilde{\chi}}^2},
\end{eqnarray}
Then Eq.(14) becomes

\begin{eqnarray}\label{eq22}
z(1-z)w^{\prime\prime}(z)+\left[\tilde{c}-\left(\tilde{a}+\tilde{b}+1\right)z \right]w^\prime(z)-\tilde{a}\tilde{b}w(z)=0, 
\end{eqnarray}
where

\begin{eqnarray}\label{eq23}
\tilde{\eta}=\frac{1}{2}\sqrt{1 + 4\left(\tilde{\gamma}D_2+\tilde{\mu}\right) a^2},
\end{eqnarray}

\begin{eqnarray}\label{eq24}
\tilde{a}=\tilde{\sigma}+\tilde{\tau}+\frac{1}{2}-\tilde{\eta},
\end{eqnarray}

\begin{eqnarray}\label{eq25}
\tilde{b}=\tilde{\sigma}+\tilde{\tau}+\frac{1}{2}+\tilde{\eta},
\end{eqnarray}

\begin{eqnarray}\label{eq26}
\tilde{c}=2\tilde{\sigma}+1,
\end{eqnarray}
The complete solution to Eq.(22) is

\begin{fleqn}
\begin{equation}\label{eq27}
\begin{aligned}
w(z)=&\ C_1 \ {}_2F_1 \left(\tilde{\sigma}+\tilde{\tau}-\tilde{\eta}+\frac{1}{2},\ \tilde{\sigma}+\tilde{\tau}+\tilde{\eta}+\frac{1}{2}, \ 2\tilde{\sigma}+1;\ z   \right) \\
&+ C_2 z^{-2\tilde{\mu}} {}_2F_1\left(-\tilde{\sigma}+\tilde{\tau}-\tilde{\eta}+\frac{1}{2}, \ -\tilde{\sigma}+\tilde{\tau}+\tilde{\eta}+\frac{1}{2}, \ -2\tilde{\sigma}+1; \ z    \right).
  \end{aligned}
\end{equation}
\end{fleqn}
For bound states $\mathcal{E}<M$, and by using boundary condition $z\rightarrow 0 \ (r\rightarrow \infty),\ G_{n\kappa}(z) \rightarrow 0$, the allowed solution is

\begin{eqnarray}\label{eq28}
G_{n\kappa}(z)=z^{\tilde{\sigma}}(1-z)^{\tilde{\tau}}  {}_2F_1 \left(\tilde{\sigma}+\tilde{\tau}-\tilde{\eta}+\frac{1}{2}, \ \tilde{\sigma}+\tilde{\tau}+\tilde{\eta}+\frac{1}{2}, \ 2\tilde{\sigma}+1; \ z   \right). 
\end{eqnarray}
Considering another boundary condition near the origin $z\rightarrow 1 \ (r\rightarrow 0)$ and applied to Eq.(28), yields

\begin{equation}\label{eq29}
\begin{aligned}
G_{n\kappa}(z) \sim &\ (1-z)^{\tilde{\tau}} \frac{\Gamma(2\tilde{\sigma}+1)\Gamma(-2\tilde{\tau})}
{\Gamma\left(\tilde{\sigma}-\tilde{\tau}+\tilde{\eta}+\frac{1}{2} \right)  
\Gamma\left(\tilde{\sigma}-\tilde{\tau}-\tilde{\eta}+\frac{1}{2} \right) } \\
&+(1-z)^{-\tilde{\tau}} \frac{\Gamma(2\tilde{\sigma}+1)\Gamma(2\tilde{\tau})}
{\Gamma\left(\tilde{\sigma}+\tilde{\tau}-\tilde{\eta}+\frac{1}{2} \right)  
\Gamma\left(\tilde{\sigma}+\tilde{\tau}+\tilde{\eta}+\frac{1}{2} \right) }.
\end{aligned}
\end{equation}

\noindent
By applying the boundary condition $G_{n\kappa}(1)=0$ in the neighborhood of $r=0$ and using an approximation $1-z=e^{-R/a}$ for realistic nuclei, we get

\begin{eqnarray}\label{eq30}
\frac{\Gamma(2\tilde{\tau})\Gamma  \left(\tilde{\sigma}+\tilde{\eta}^\prime+1-\tilde{\tau} \right)  \Gamma  \left(\tilde{\sigma}-\tilde{\eta}^\prime-\tilde{\tau} \right)}
{\Gamma(-2\tilde{\tau})\Gamma  \left(\tilde{\sigma}+\tilde{\eta}^\prime+1+\tilde{\tau} \right)  \Gamma  \left(\tilde{\sigma}-\tilde{\eta}^\prime+\tilde{\tau} \right)}
e^{\frac{2\tilde{\tau}R}{a}}=-1,
\end{eqnarray}
where $\tilde{\eta}^\prime=\tilde{\eta}-1/2$. We note that $\tilde{\sigma}^2-\tilde{\chi}^2<0$, the parameter $\tilde{\tau}$ becomes imaginary

\begin{eqnarray}\label{eq31}
\tilde{\tau}=i\lambda,  \hspace{2 em} \lambda=\sqrt{\tilde{\chi}^2-\tilde{\sigma}^2},
\end{eqnarray}
Therefore, from Eq.(30), we can get the energy eigenvalues equation in the case of exact pseudospin symmetry as follow

\begin{eqnarray}\label{eq32}
\tilde{\xi}-\tilde{\zeta}-\tilde{\omega}+\frac{\lambda R}{a}=\left(n+\frac{1}{2} \right)\pi; \hspace{2em} n = 0,\pm 1,\pm 2,\pm 3, \ldots 
\end{eqnarray}
where

\begin{eqnarray}\label{eq33}
\tilde{\xi}=\mathrm{arg}\ \Gamma(2i\lambda), \ \ \tilde{\zeta}=\mathrm{arg}\ \Gamma\left(\tilde{\sigma}+\tilde{\eta}^\prime+1+i\lambda \right), \ \ \tilde{\omega}=\mathrm{arg}\ \Gamma\left(\tilde{\sigma}-\tilde{\eta}^\prime+i\lambda \right),  
\end{eqnarray}
and corresponding the lower spinor component

\begin{eqnarray}\label{eq34}
G_{n\kappa}(r)=\tilde{N}\left(1+e^\frac{r-R}{a} \right)^{-(\tilde{\sigma}+i\lambda)} e^\frac{i\lambda(r-R)}{a} {}_2F_1\left(\tilde{a},\ \tilde{b},\ \tilde{c};\ \frac{1}{1+e^\frac{r-R}{a}} \right). 
\end{eqnarray}
We also obtain the upper spinor component by using Eq.(4) as follow

\begin{fleqn}
\begin{equation}\label{eq35}
\begin{split}
F_{n\kappa}(r)=& \ \frac{\tilde{N}}{M-\mathcal{E}+C_{ps}} \left(1+e^\frac{r-R}{a} \right)^{-(\tilde{\sigma}+i\lambda)} e^\frac{i\lambda(r-R)}{a}  
\\ 
& \times \left\{  \left[-\frac{\kappa+H}{r}+\frac{i\lambda}{a}-\frac{\tilde{\sigma}+i\lambda}{a\left(1+e^\frac{-(r-R)}{a} \right) } \right]  
{}_2F_1\left(\tilde{a}, \ \tilde{b}, \ \tilde{c}; \ \frac{1}{1+e^\frac{r-R}{a}} \right) 
\right. \\
& \hspace{2em} \left.
+\ \frac{\tilde{a}\tilde{b}}{\tilde{c}} {}_2F_1\left(\tilde{a}+1,\ \tilde{b}+1, \ \tilde{c}+1;\ \frac{1}{1+e^\frac{r-R}{a}} \right)
\vphantom{\frac{\tilde{\sigma}+i\lambda}{a\left(1+e^\frac{-(r-R)}{a} \right) }} 
\right\}.
\end{split}
\end{equation}
\end{fleqn}
where $F_{n\kappa}$ is admissible for $\mathcal{E}\neq M+C_{ps}$ and only for bound negative energy states solutions \cite{ginocchio2005relativistic}. 
Moreover, we can reduce the energy eigenvalues in Eq.(32) into Schrödinger equation by using these transfomations $\mathcal{E}+M-C_s \rightarrow 2\mu^\prime,\ (\kappa+H)(\kappa+H+1)\rightarrow l^\prime(l^\prime+1),\ \mathcal{E}-M \rightarrow E_{NR}$ and if we take $\mathit{\Delta}_1=0$, then Eq.(32) becomes

\begin{fleqn}
\begin{equation}\label{eq36}
\begin{split}
&\mathrm{arg} \  \Gamma \left(ia\sqrt{8\mu^\prime(E_{NR}+\mathit{\Delta}_0)- 4l^\prime(l^\prime+1)(D_0+D_1+D_2)/R^2}\right)
\\
&- \mathrm{arg}  \ \Gamma \left( a\sqrt{l^\prime(l^\prime+1)D_0/R^2-2\mu^\prime E_{NR}}+\sqrt{l^\prime(l^\prime+1)D_2a^2/R^2+1/4}
\right.
\\ 
& \hspace{4em} \left. +\ 1/2 +   ia\sqrt{2\mu^\prime (E_{NR}+\mathit{\Delta}_0)-l^\prime(l^\prime+1)(D_0+D_1+D_2)/R^2} \right)
\\
& - \mathrm{arg}  \ \Gamma  \left(a\sqrt{l^\prime(l^\prime+1)D_0/R^2-2\mu^\prime E_{NR}}-\sqrt{l^\prime(l^\prime+1)D_2a^2/R^2+1/4}
\right.
\\ 
& \hspace{4em} \left.  +\ 1/2 +   ia\sqrt{2\mu^\prime (E_{NR}+\mathit{\Delta}_0)-l^\prime(l^\prime+1)(D_0+D_1+D_2)/R^2} \right)
\\
&+\sqrt{2\mu^\prime(E_{NR}+\mathit{\Delta}_0)R^2-l^\prime(l^\prime+1)(D_0+D_1+D_2)}=(2n^\prime-1)\frac{\pi}{2}.
\end{split}
\end{equation}
\end{fleqn}

\noindent
where $E_{NR}$, $\mu^\prime$, $l^\prime$, and $n^\prime$ are the non-relativistic energy, reduced mass, the azimuthal quantum number, and the principal quantum number, respectively. The result in Eq.(36) consistent with expression (38) of ref. \cite{pahlavani2012study} if we take the spin orbit parameter $V_{LS}^{(0)}=0$.

\begin{table}[H]
\captionsetup{width=1.5\textwidth}
\caption*{Table 1. The bound states energy eigenvalues in the unit of MeV in the case of exact pseudospin symmetry}
\label{tab:tabel1}
\centerline{
\begin{tabular}{l l l l l l l l l} 
 \hline
 $\tilde{l}$ & $n, \kappa <0$ & $(l,j)$ & $\tilde{E}_{n,\kappa <0(H=0)}$ & $\tilde{E}_{n,\kappa <0(H=-1.5)}$ & $n-1, \kappa >0$ & $(l+2,j+1)$ & $\tilde{E}_{n-1,\kappa >0(H=0)}$  & $\tilde{E}_{n-1,\kappa >0(H=-1.5)}$ 
  \\ [0.5ex]
 \hline 
$0$ & $-$ & $-$ & $-$ & $-$ & $0,1$	& $0p_{1/2}$ & $-42.8961$ & $-42.8335$\\
$0$ & $-$ & $-$ & $-$ & $-$ & $1,1$	& $1p_{1/2}$ & $-31.8026$ & $-31.7044$\\
$0$ & $-$ & $-$ & $-$ & $-$ & $2,1$	& $2p_{1/2}$ & $-16.7022$ & $-16.5811$\\
$1$ & $1,-1$ & $1s_{1/2}$ & $-42.7642$ & $-40.3580$ & $0,2$ & $0d_{3/2}$ & $-42.7642$ & $-42.9207$\\
$1$ & $2,-1$ & $2s_{1/2}$ & $-31.5396$ & $-28.8446$ & $1,2$ & $1d_{3/2}$ & $-31.5396$ & $-31.8355$\\
$1$ & $3,-1$ & $3s_{1/2}$ & $-16.3727$ & $-15.0351$ & $2,2$ & $2d_{3/2}$ & $-16.3727$ & $-16.7421$\\
$2$ & $1,-2$ & $1p_{3/2}$ & $-41.5158$ & $-37.4032$ & $0,3$ & $0f_{5/2}$ & $-41.5158$ & $-42.8335$\\
$2$ & $2,-2$ & $2p_{3/2}$ & $-30.0065$ & $-25.8873$ & $1,3$ & $1f_{5/2}$ & $-30.0065$ & $-31.7044$\\
$2$ & $3,-2$ & $3p_{3/2}$ & $-15.6299$ & $-7.2731$ & $2,3$ &  $2f_{5/2}$ & $-15.6299$ & $-16.5811$\\
$3$ & $1,-3$ & $1d_{5/2}$ & $-38.9874$ & $-33.5910$ & $0,4$ & $0g_{7/2}$ & $-38.9874$ & $-42.4616$\\
$3$ & $2,-3$ & $2d_{5/2}$ & $-27.4716$ & $-22.0858$ & $1,4$ & $1g_{7/2}$ & $-27.4716$ & $-31.2969$\\
$3$ & $3,-3$ & $3d_{5/2}$ & $-8.8563$ & $-3.4773$ & $2,4$ &   $2g_{7/2}$ & $-8.8563$ & $-16.0637$\\ [0.5ex]
 \hline
\end{tabular}}
\end{table}

In order to present an illustrative example for the energy eigenvalue equation, we use the following parameters for numerical calculation: $M = 939.5654$ MeV/$c^2$, $a = 0.65$ fm, $R = 7$ fm, $\mathit{\Delta}_0 = 750$ MeV, $\mathit{\Delta}_1= 50$ MeV, $C_{ps} = -50$ MeV, and $\hbar c = 197.3269$ MeV fm. Moreover, the binding energy for a bound Dirac particle in the case of exact pseudospin symmetry is given by $\tilde{E}=\mathcal{E}-M$. From Table 1, one can observe that the energies of all nucleon states increase with increases in quantum numbers $n$ or $\tilde{l}$. Furthermore, pseudospin doublets have energy degeneracy in the absence of the tensor potential $(H=0)$, but in the case $H\neq 0$, the energy degeneracy between pseudospin doublets is removed. Therefore, we can say that the tensor interactions remove energy degeneracy for pseudospin doublets, and this present result agrees with the previous one \cite{aydougdu2010pseudospin, akcay2009dirac, akcay2009exact, alberto2005tensor, lisboa2004pseudospin}. On the other hand, we can see from Table 1. that the tensor interaction $(H \neq 0)$ make energies of aligned and unaligned pseudospin move in opposite direction. This can be explained by considering energy eigenvalue in Eq.(32); the energy eigenvalues depend on pseudospin dependent term $2\kappa H$ in $\tilde{\gamma}$. The factor $\kappa$ takes negative and positive values depending on the pseudospin alignment. Then, we also observe that the parameter of tensor potential for $H<0$ provide higher energies for the pseudospin aligned states and lower energies for the pseudospin unaligned states. As a result, we get a positive value for pseudospin energy splitting $\left(\Delta\tilde{E}=\tilde{E}_{\tilde{l}j=\tilde{l}-1/2}-\tilde{E}_{\tilde{l}j=\tilde{l}+1/2}\right)$.

\begin{figure}[H]
\centering
\includegraphics{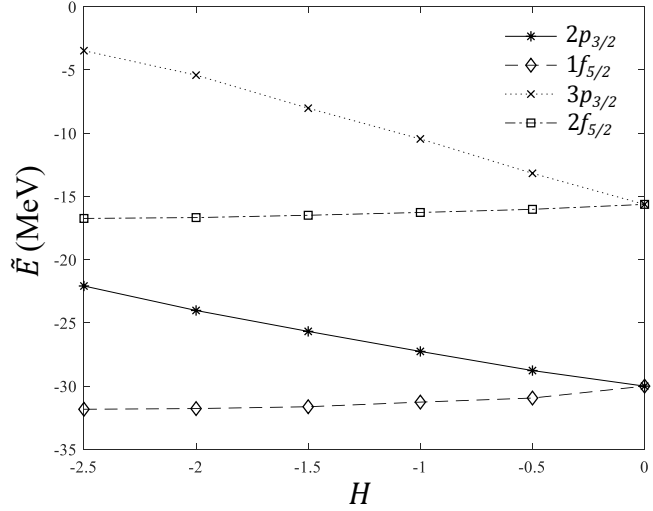}
\captionsetup{width=1\textwidth}
\caption*{Figure 1. Pseudospin energy eigenvalues as a function of $H$ for the 
pseudospin doublets $(2p_{3/2},1f_{5/2})$ and $(3p_{3/2},2f_{5/2})$
}
\label{tab:figure1}
\end{figure}

\indent
We investigate the behavior of pseudospin energy eigenvalues in the presence of tensor potential in Figure 1. for pseudospin doublets $(2p_{3/2},1f{_5/2})$ and $(3p_{3/2},2f_{5/2})$. From Figure 1, the energy splittings of pseudospin doublets for $(3p_{3/2},2f_{5/2})$ have a greater value than $(2p_{3/2},1f_{5/2})$ in the case of $H\neq 0$. Furthermore, this energy splitting of pseudospin doublets increases while $H$ decreases. It can be understood again by the dependence of energy eigenvalues in Eq.(32) with pseudospin dependent term $2\kappa H$ in $\tilde{\gamma}$. So even though the energies of pseudospin aligned states decrease as $H$ increases, the energies of the pseudospin unaligned states can still increase with increasing $H$.
\\
\\
\begin{figure}[H]
\centering
\includegraphics{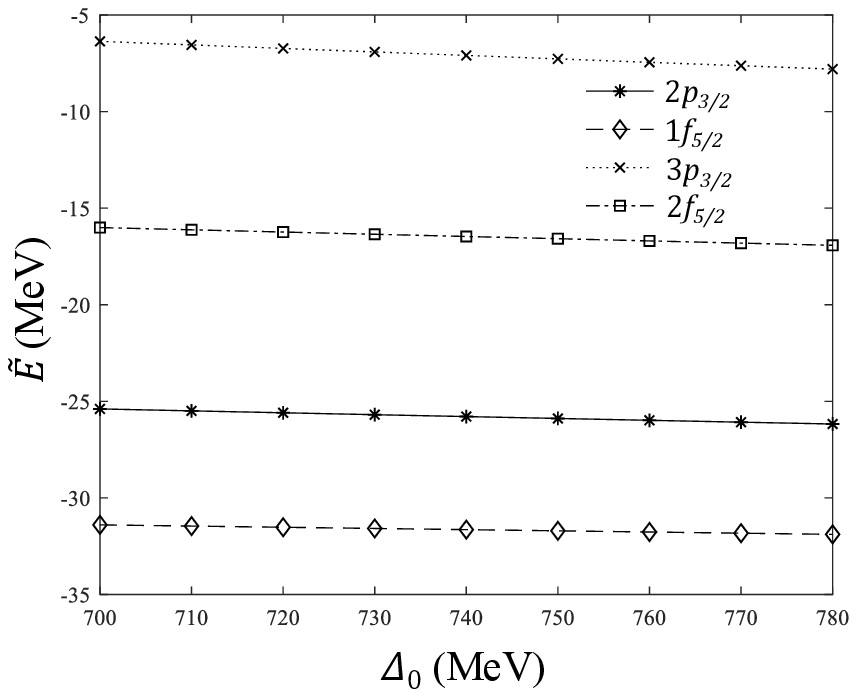}
\captionsetup{width=1\textwidth}
\caption*{Figure 2. Pseudospin energy eigenvalues as a function of $\mathit{\Delta}_0$ for the 
pseudospin doublets $(2p_{3/2},1f_{5/2})$ and $(3p_{3/2},2f_{5/2})$
}
\label{tab:figure2}
\end{figure}

\begin{figure}[H]
\centering
\includegraphics{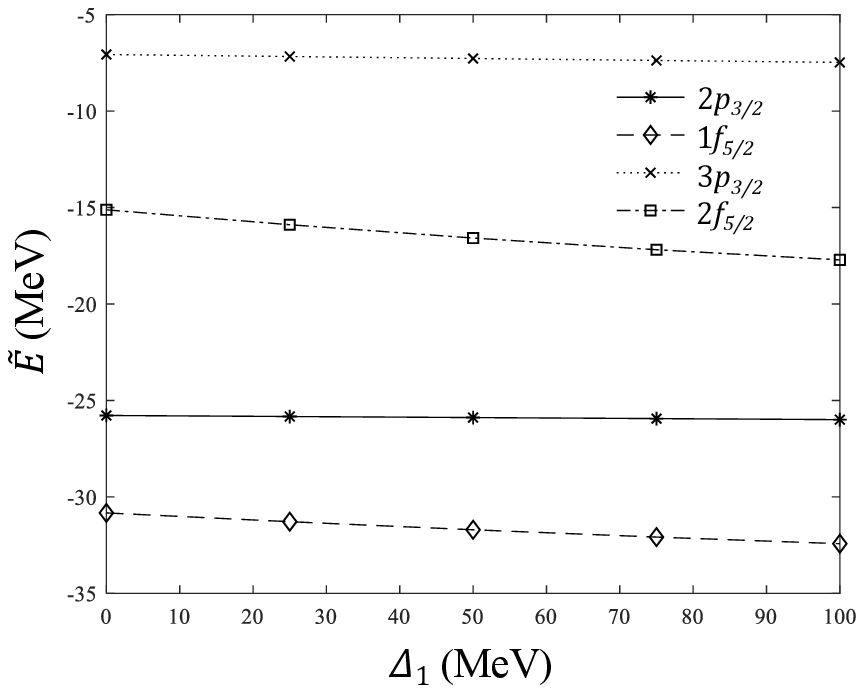}
\captionsetup{width=1\textwidth}
\caption*{Figure 3. Pseudospin energy eigenvalues as a function of $\mathit{\Delta}_1$ for the 
pseudospin doublets $(2p_{3/2},1f_{5/2})$ and $(3p_{3/2},2f_{5/2})$
}
\label{tab:figure3}
\end{figure}

\indent
The sensitivity of the pseudospin energy eigenvalues to the width of potential well $\mathit{\Delta}_0$ and $\mathit{\Delta}_1$ for $H=-1.5$ are given in Figures 2 and 3, respectively. It can be noted that when we vary the width of potential well $\mathit{\Delta}_0$ or $\mathit{\Delta}_1$ without the presence of tensor potential, there is no energy splitting that occurs between pseudospin doublets. From Figure 2, we observe that both of the pseudospin energy splitting and eigenvalues for all of the pseudospin doublets have a slight decrease when $\mathit{\Delta}_0$ increases. Meanwhile, in Figure 3, the energy eigenvalues for $1f_{5/2}$ and $2f_{5/2}$ states have a slight decrease to the vary of $\mathit{\Delta}_1$ than $2p_{3/2}$ and $3p_{3/2}$ states, but the energy splitting for all of the pseudospin doublets have a slight increase as $\mathit{\Delta}_1$ increases. Therefore, from Figures 2 and 3, both of the pseudospin energy splitting and eigenvalues in the presence of tensor potential are insensitive to the changes of $\mathit{\Delta}_0$ or $\mathit{\Delta}_1$.

\section{The spin symmetry}
Now, we consider in the case of exact spin symmetry ($d\mathit{\Delta}(r)/dr=0,$ i.e., $\mathit{\Delta}(r)=C_s=$ const). Assuming $\mathit{\Sigma}(r)$ and $U(r)$ are potential of the generalized Woods-Saxon type and potential of the Coulomb-like tensor, respectively, defined as follow

\begin{eqnarray}\label{eq37}
\mathit{\Delta(r)}=C_{s},\hspace{2em} \mathit{\Sigma(r)}=-\frac{\mathit{\Sigma}_0}{1+e^{\frac{r-R}{a}}}-\frac{\mathit{\Sigma}_1 e^{\frac{r-R}{a}}}{\left(1+e^{\frac{r-R}{a}} \right)^2 }, \hspace{2em}  U(r)=-\frac{H}{r},
\end{eqnarray}

\noindent
where $\mathit{\Sigma}_0$ and $\mathit{\Sigma}_1$ represent the depth of the potential well. The parameters $a$, $R$, and $H$ are the diffusivity of nuclear surface, the width of the potential well, and constants, respectively. Then, Eq.(5) is turned into the following form

\begin{fleqn}
\begin{equation}\label{eq38}
\begin{split}
&  \left[
\vphantom{\frac{\mathit{\Sigma}_1e^\frac{r-R}{a}}{\left(1+e^\frac{r-R}{a} \right)^2 }}
    \frac{d^2}{dr^2}-\frac{(\kappa+H)(\kappa+H+1)}{r^2} 
\right.\\
&  \left.
\hspace{1em}+\ (\mathcal{E}+M-C_s)\left(\mathcal{E}-M+\frac{\mathit{\Sigma}_0}{1+e^\frac{r-R}{a}}+\frac{\mathit{\Sigma}_1e^\frac{r-R}{a}}{\left(1+e^\frac{r-R}{a} \right)^2 } \right)       \right]F_{n\kappa}(r)=0. 
\end{split}
\end{equation}
\end{fleqn}

\noindent
This Equation has the same form as the Eq.(8), so we applied the Pekeris approximation using the same notations in Eq.(9) and introduce similar variable $z=\left(1+e^\frac{r-R}{a}\right)^{-1}$ like in the previous sections, we obtain

\begin{fleqn}
\begin{equation}\label{eq39}
\begin{split}
& \left[\frac{d^2}{dz^2} + \frac{(1-2z)}{z(1-z)}\frac{d}{dz} 
+ \frac{1}{z^2(1-z)^2} \bigl(-\left(\gamma D_2 + \mu \right)a^2z^2 + \left(\varpi + \mu-\gamma D_1 \right)a^2z 
\right. \\ 
& \left.
\hspace{14.5em} \ - \left(\gamma D_0-\mathit{\Lambda} \right)a^2    
\vphantom{\frac{d^2}{dz^2}}  
\bigr)   
 \right]  F_{n\kappa}(z)=0.
\end{split}
\end{equation}
\end{fleqn}

\noindent
where

\begin{eqnarray}\label{eq40}
\gamma=\frac{(\kappa+H)(\kappa+H+1)}{R^2},
\end{eqnarray}

\begin{eqnarray}\label{eq41}
\varpi=(\mathcal{E}+M-C_s)\mathit{\Sigma}_0,
\end{eqnarray}

\begin{eqnarray}\label{eq42}
\mathit{\Lambda}=(\mathcal{E}-M)(\mathcal{E}+M-C_s),
\end{eqnarray}

\begin{eqnarray}\label{eq43}
\mu=(\mathcal{E}+M-C_s)\mathit{\Sigma}_1,
\end{eqnarray}

\noindent
Eq.(39) have the same form as Eq.(14). Thus, we applied the same procedure as that used in the previous section. So, the corresponding upper spinor component is

\begin{eqnarray}\label{eq44}
F_{n\kappa}=N \left(1+e^\frac{r-R}{a} \right)^{-(\sigma+i\delta)} e^\frac{i\delta(r-R)}{a} {}_2F_1\left(a^\prime,\ b^\prime,\ c^\prime; \frac{1}{1+e^\frac{r-R}{a}} \right), 
\end{eqnarray}

\noindent
with

\begin{eqnarray}\label{eq45}
\sigma=\sqrt{(\gamma D_0-\mathit{\Lambda})a^2},
\end{eqnarray}

\begin{eqnarray}\label{eq46}
\chi=\sqrt{\bigl(\varpi-\gamma(D_1+D_2)\bigr)a^2},
\end{eqnarray}

\begin{eqnarray}\label{eq47}
\tau=i\delta, \hspace{2em}  \delta=\sqrt{\chi^2-\sigma^2},
\end{eqnarray}

\noindent
and

\begin{eqnarray}\label{eq48}
\eta=\frac{1}{2}\sqrt{1+4(\gamma D_2+\mu)a^2},
\end{eqnarray}

\begin{eqnarray}\label{eq49}
a^\prime=\sigma+\tau+\frac{1}{2}-\eta,
\end{eqnarray}

\begin{eqnarray}\label{eq50}
b^\prime=\sigma+\tau+\frac{1}{2}+\eta,
\end{eqnarray}

\begin{eqnarray}\label{eq51}
c^\prime=2\sigma+1,
\end{eqnarray}

\noindent
with Eq.(3), the lower spinor component is obtained as

\begin{fleqn}
\begin{equation}\label{eq52}
\begin{split}
G_{n\kappa}(r)=& \ \frac{N}{M+\mathcal{E}-C_{s}} \left(1+e^\frac{r-R}{a} \right)^{-(\sigma+i\delta)} e^\frac{i\delta(r-R)}{a}  
\\ 
& \times \left\{  \left[\frac{\kappa+H}{r}+\frac{i\delta}{a}-\frac{\sigma+i\delta}{a\left(1+e^\frac{-(r-R)}{a} \right) } \right]  
{}_2F_1\left(a^\prime, \ b^\prime, \ c^\prime; \ \frac{1}{1+e^\frac{r-R}{a}} \right) 
\right. \\
& \hspace{2em} \left.
+\ \frac{a^\prime b^\prime}{c^\prime} {}_2F_1\left(a^\prime+1,\ b^\prime+1, \ c^\prime+1;\ \frac{1}{1+e^\frac{r-R}{a}} \right)
\vphantom{\frac{\tilde{\sigma}+i\lambda}{a\left(1+e^\frac{-(r-R)}{a} \right) }} 
\right\}.
\end{split}
\end{equation}
\end{fleqn}

\noindent
where $G_{n\kappa}$ is admissible for $\mathcal{E}\neq -M+C_s$ and only for bound positive energy states solutions \cite{ginocchio2005relativistic}.

\indent
The energy eigenvalues equation in the case of exact spin symmetry can also be obtained as follow

\begin{eqnarray}\label{eq53}
\xi-\zeta-\omega+\frac{\delta R}{a}=\left(n+\frac{1}{2} \right)\pi; \hspace{2em} n = 0,\pm 1,\pm 2,\pm 3, \ldots 
\end{eqnarray}

\noindent
where

\begin{eqnarray}\label{eq54}
\xi=\mathrm{arg}\ \Gamma(2i\delta), \ \ \zeta=\mathrm{arg}\ \Gamma\left(\sigma+\eta^\prime+1+i\delta \right), \ \ \omega=\mathrm{arg}\ \Gamma\left(\sigma-\eta^\prime+i\delta \right),  
\end{eqnarray}

\noindent
if we take $l=H=\mathit{\Sigma}_1=0$ in Eq.(53), then Eq.(53) becomes

\begin{equation}\label{eq55}
\begin{split}
&\mathrm{arg} \ \Gamma\left(2ia\sqrt{(M+\mathcal{E}_{n,-1}-C_s)(\mathcal{E}_{n,-1}-M+\mathit{\Sigma}_0)}\right) \\
&- 2\ \mathrm{arg} \ \Gamma \left(a\sqrt{(M-\mathcal{E}_{n,-1})(M+\mathcal{E}_{n,-1}-C_s)}
\right.
\\ 
& \hspace{5em} \left.
+ \ ia\sqrt{(M+\mathcal{E}_{n,-1}-C_s)(\mathcal{E}_{n,-1}-M+\mathit{\Sigma}_0)}\right) \\
& - \mathrm{tan}^{-1}\sqrt{(\mathcal{E}_{n,-1}-M+\mathit{\Sigma}_0)/(M-\mathcal{E}_{n,-1})}\\
&+R\sqrt{(M+\mathcal{E}_{n,-1}-C_s)(\mathcal{E}_{n,-1}-M+\mathit{\Sigma}_0)}=\left(n+\frac{1}{2}\right)\pi.
\end{split}
\end{equation}

\noindent
The result in Eq.(55) consistent with expression (19) of ref. \cite{guo2005solution}. Moreover, we can reduce Eq.(53) into Schrödinger equation by using the same procedure as that used to get Eq.(36), then the transformation of Eq.(53) will be consistent with expression (38) of ref. \cite{pahlavani2012study} if we take the spin orbit parameter $V_{LS}^{(0)} = 0$.


\begin{table}[H]
\captionsetup{width=1.5\textwidth}
\caption*{Table 2. The bound states energy eigenvalues in the unit of MeV in the case of exact spin symmetry}
\label{tab:tabel2}
\centerline{
\begin{tabular}{l l l l l l l l l}
 \hline
 $l$ & $n, \kappa <0$ & $(l,j)$ & $E_{n,\kappa <0(H=0)}$ & $E_{n,\kappa <0(H=-1.5)}$ & $n, \kappa >0$ & $(l,j)$ & $E_{n,\kappa >0(H=0)}$  & $E_{n,\kappa >0(H=-1.5)}$
  \\ [0.5ex]
 \hline
$0$ & $0,-1$ & $0s_{1/2}$ & $49.1470$ & $47.7877$ & $-$ & $-$ & $-$	& $-$\\
$0$ & $1,-1$ & $1s_{1/2}$ & $42.8961$ & $42.4616$ & $-$ & $-$ & $-$	& $-$\\
$0$ & $2,-1$ & $2s_{1/2}$ & $31.8026$ & $31.2969$ & $-$ & $-$ & $-$	& $-$\\
$1$ & $0,-2$ & $0p_{3/2}$ & $48.5261$ & $45.6458$ & $0,1$ & $0p_{1/2}$ & $48.5261$ & $49.1711$\\
$1$ & $1,-2$ & $1p_{3/2}$ & $42.7642$ & $40.3580$ & $1,1$ & $1p_{1/2}$ & $42.7642$ & $42.9207$\\
$1$ & $2,-2$ & $2p_{3/2}$ & $31.5396$ & $28.8446$ & $2,1$ & $2p_{1/2}$ & $31.5396$ & $31.8355$\\
$2$ & $0,-3$ & $0d_{5/2}$ & $46.8290$ & $42.5949$ & $0,2$ & $0d_{3/2}$ & $46.8290$ & $49.0489$\\
$2$ & $1,-3$ & $1d_{5/2}$ & $41.5158$ & $37.4032$ & $1,2$ & $1d_{3/2}$ & $41.5158$ & $42.8335$\\
$2$ & $2,-3$ & $2d_{5/2}$ & $30.0065$ & $25.8873$ & $2,2$ & $2d_{3/2}$ & $30.0065$ & $31.7044$\\
$3$ & $0,-4$ & $0f_{7/2}$ & $44.2350$ & $38.6270$ & $0,3$ & $0f_{5/2}$ & $44.2350$ & $47.7877$\\
$3$ & $1,-4$ & $1f_{7/2}$ & $38.9874$ & $33.5910$ & $1,3$ & $1f_{5/2}$ & $38.9874$ & $42.4616$\\
$3$ & $2,-4$ & $2f_{7/2}$ & $27.4716$ & $22.0858$ & $2,3$ & $2f_{5/2}$ & $27.4716$ & $31.2969$\\ [0.5ex]
\hline
\end{tabular}}
\end{table}

\indent
We present the example of numerical calculation using the following parameters: $M = 939.5654$ MeV/$c^2$, $a = 0.65$ fm, $R = 7$ fm, $\mathit{\Sigma}_0 = -750$ MeV, $\mathit{\Sigma}_1 = -50$ MeV, $C_s = 50$ MeV, and $\hbar c = 197.3269$ MeV fm. Besides that, the binding energy in the case of exact spin symmetry is given by $E=\mathcal{E}+M$. From Table 2, one can observe that the energies of all nucleon states decrease with increases in quantum numbers $n$ or $l$. Also, one can see that spin doublets have energy degeneracy in the absence of the tensor potential $(H=0)$, but this degeneracy is removed in the case of $H\neq 0$. Our results show that the energy degeneracy for spin doublets is removed by tensor interaction. This present result is in good agreement with those obtained previously \cite{aydougdu2010pseudospin, akcay2009dirac, akcay2009exact, alberto2005tensor, lisboa2004pseudospin}. Like in the case of exact pseudospin symmetry, we also found that the energies of the aligned and unaligned spin move in the opposite direction in the presence of tensor potential. This happens because the energy eigenvalues in Eq.(53) depends on the term $2\kappa H$ through $\gamma$. The spin-dependent term $2\kappa H$ takes negative and positives values depending on the values of $\kappa$, respectively. Then, we can see that the choice for $H<0$ provide lower energies for the spin aligned states and higher energies for the spin unaligned states. Thus, we get a positive value for spin energy splitting $(\Delta E=E_{lj=l-1/2}-E_{lj=l+1/2})$.

\begin{figure}[H]
\centering
\includegraphics{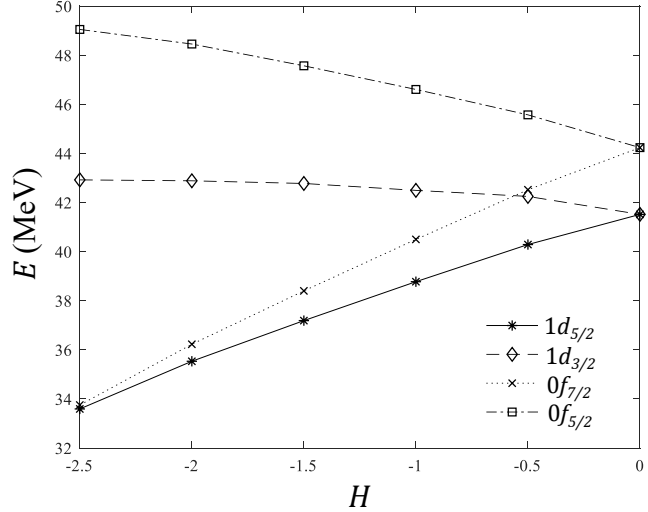}
\captionsetup{width=1\textwidth}
\caption*{Figure 4. Spin energy eigenvalues as a function of $H$ for the spin doublets $(1d_{5/2},1d_{3/2})$ and $(0f_{7/2},0f_{5/2})$
}
\label{tab:figure4}
\end{figure}

\indent
We investigate how the spin energy eigenvalues dependent on the parameters $H$, $\mathit{\Sigma}_0$ and $\mathit{\Sigma}_1$ in Figures 4, 5, and 6, respectively. We consider the spin doublets $(1d_{5/2},1d_{3/2})$ and $(0f_{7/2},0f_{5/2})$ as an example. From Figure 4, the energy splitting of spin doublets $(0f_{7/2},0f_{5/2})$ have a greater value than $(1d_{5/2},1d_{3/2})$ in the case of $H\neq 0$. Like in the case of exact pseudospin symmetry, the energy splitting of spin doublets increases while $H$ decreases. This is because of the dependence of energy eigenvalues in Eq.(53) on the spin-dependent term $2\kappa H$ in $\gamma$.
\\
\\
\\
\\
\\
\\
\begin{figure}[H]
\centering
\includegraphics{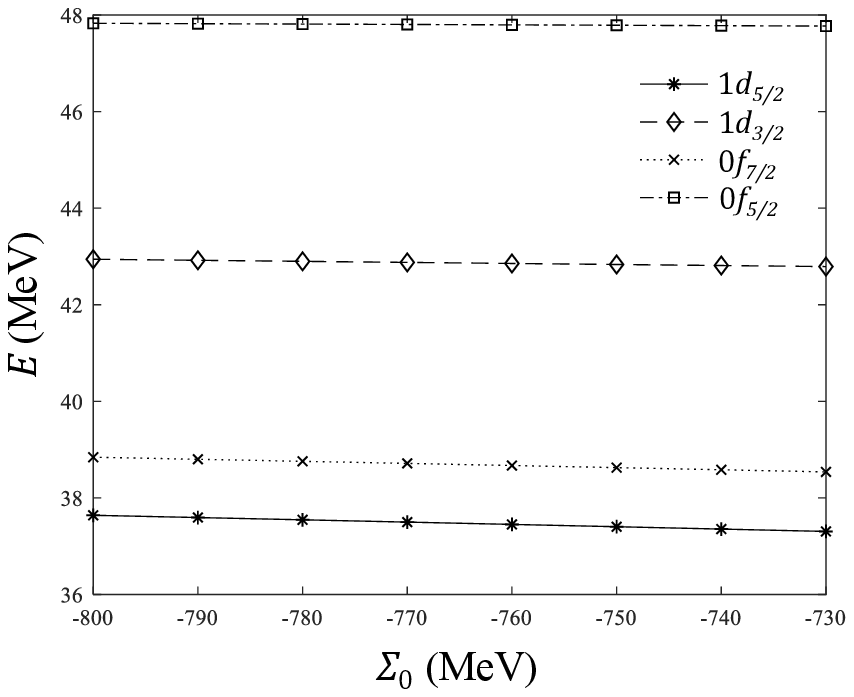}
\captionsetup{width=1\textwidth}
\caption*{Figure 5. Spin energy eigenvalues as a function of $\mathit{\Sigma}_0$ for the spin doublets $(1d_{5/2},1d_{3/2})$ and $(0f_{7/2},0f_{5/2})$
}
\label{tab:figure5}
\end{figure}

\begin{figure}[H]
\centering
\includegraphics{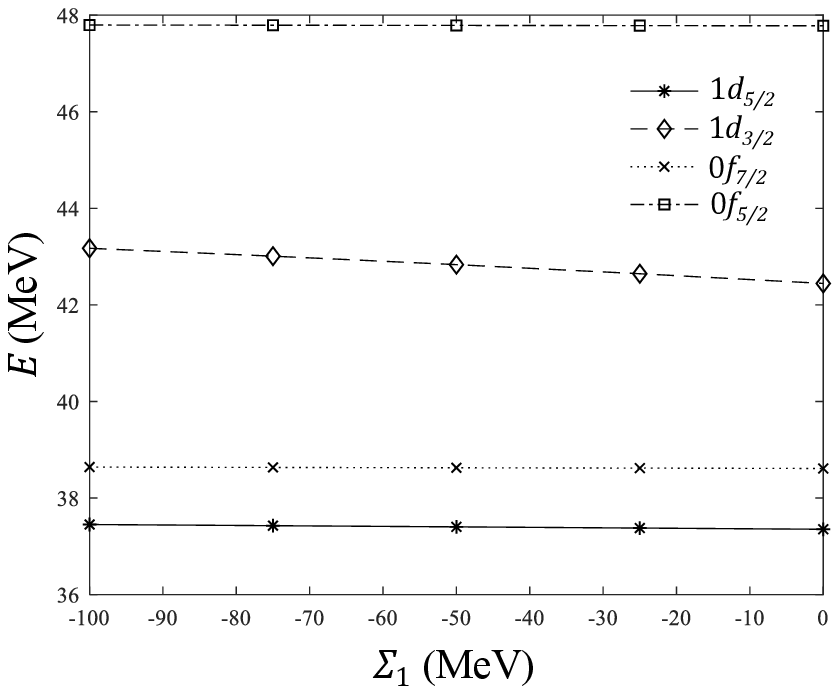}
\captionsetup{width=1\textwidth}
\caption*{Figure 6. Spin energy eigenvalues as a function of $\mathit{\Sigma}_1$ for the spin doublets $(1d_{5/2},1d_{3/2})$ and $(0f_{7/2},0f_{5/2})$
}
\label{tab:figure6}
\end{figure}

\indent
In Figures 5 and 6, we vary the width of potential well $\mathit{\Sigma}_0$ and $\mathit{\Sigma}_1$ in the presence of tensor potential for $H=-1.5$, respectively, to investigate the sensitivity of spin energy splitting and eigenvalues. We observe that there is no energy splitting occurs between spin doublets when we vary the width of the potential well $\mathit{\Sigma}_0$ and $\mathit{\Sigma}_1$ without the presence of tensor potential. From Figure 5, one can observe that the spin energy splitting have a slight increase with increasing $\mathit{\Sigma}_0$, while the energy eigenvalues have a slight decrease with increasing $\mathit{\Sigma}_0$. Then, in Figure 6, the energy eigenvalues for $1d_{3/2}$ state have a slight decrease to the vary of $\mathit{\Sigma}_1$ than the other states. Like in the case of exact pseudospin symmetry, from Figures 5 and 6, we found that both of the spin energy splitting and eigenvalues in the presence of tensor potential are insensitive to the changes of $\mathit{\Sigma}_0$ or $\mathit{\Sigma}_1$.

\section{Conclusions}
In this work, we have obtained approximate analytical solutions of the Dirac equation for the generalized Woods-Saxon potential including Coulomb-like tensor potential in the case of exact pseudospin and spin symmetry. By carefully examining the asymptotic behavior of the wave function in neighborhood $r=R$, the bound states energy eigenvalues and corresponding radial wavefunctions have been obtained in the case of exact pseudospin and spin symmetry limits. We have found that the tensor interactions $(H\neq 0)$ remove energy degeneracy for pseudospin (spin) doublets. These present results are in good agreement with those obtained previously \cite{aydougdu2010pseudospin, akcay2009dirac, akcay2009exact, alberto2005tensor, lisboa2004pseudospin}. We have also found that the energies of aligned and unaligned pseudospin (spin) move in the opposite direction due to the pseudospin (spin) dependent term in the energy eigenvalues equation. We have investigated the dependence of pseudospin (spin) energy eigenvalues and splitting on the parameters $H,\mathit{\Delta}_0 (\mathit{\Sigma}_0 )$ and $\mathit{\Delta}_1 (\mathit{\Sigma}_1 )$. In both of pseudospin (spin) doublets, the energy splitting increases while $H$ decreases. Furthermore, in the presence of tensor potential, the pseudospin (spin) energy splitting and eigenvalues are insensitive to the changes of $\mathit{\Delta}_0(\mathit{\Sigma}_0)$ or $\mathit{\Delta}_1(\mathit{\Sigma}_1)$.
\\

\section*{Acknowledgments}
This research was partly supported by the Mandatory Research Grant of Sebelas Maret University with contract number 452/UN27.21/PN/2020.
\\

\bibliographystyle{elsarticle-num}
\bibliography{References}

\end{document}